\documentclass[12pt, onecolumn, superscriptaddress,notitlepage]{revtex4-1}
\usepackage{color}
\usepackage{amsmath,amssymb,graphicx,bm,float}

\newcommand{\ket}[1]{|{#1}\rangle}

\begin{document}
\title{Spectral characterization of photon-pair sources via classical sum-frequency generation}
\author{Fumihiro Kaneda}
\affiliation{Frontier Research Institute for Interdiciplinary Sciences, Tohoku University, 6-3 Aramaki aza Aoba, Aoba-ku, Sendai 980-8578, Japan}
\affiliation{Research Institute of Electrical Communication, Tohoku University, 2-1-1 Katahira, Sendai, 980-8577, Japan}
\author{Jo Oikawa}
\affiliation{Research Institute of Electrical Communication, Tohoku University, 2-1-1 Katahira, Sendai, 980-8577, Japan}
\author{Masahiro Yabuno}
\affiliation{Advanced ICT Research Institute, National Institute of Information and Communications Technology, 588-2 Iwaoka, Nishi-ku, Kobe, Hyogo 651-2492, Japan}
\author{Fumihiro China}
\affiliation{Advanced ICT Research Institute, National Institute of Information and Communications Technology, 588-2 Iwaoka, Nishi-ku, Kobe, Hyogo 651-2492, Japan}
\author{Shigehito Miki}
\affiliation{Advanced ICT Research Institute, National Institute of Information and Communications Technology, 588-2 Iwaoka, Nishi-ku, Kobe, Hyogo 651-2492, Japan}
\affiliation{Graduate School of Engineering, Kobe University, 1-1 Rokko-dai cho, Nada-ku, Kobe 657-0013, Japan }
\author{Hirotaka Terai}
\affiliation{Advanced ICT Research Institute, National Institute of Information and Communications Technology, 588-2 Iwaoka, Nishi-ku, Kobe, Hyogo 651-2492, Japan}
\author{Yasuyoshi Mitsumori}
\affiliation{Research Institute of Electrical Communication, Tohoku University, 2-1-1 Katahira, Sendai, 980-8577, Japan}
\author{Keiichi Edamatsu}
\affiliation{Research Institute of Electrical Communication, Tohoku University, 2-1-1 Katahira, Sendai, 980-8577, Japan}

\begin{abstract}
Tailoring spectral properties of photon pairs is of great importance for optical quantum information and measurement applications. 
High-resolution spectral measurement is a key technique for engineering spectral properties of photons, making them ideal for various quantum applications. 
Here we demonstrate spectral measurements and optimization of frequency-entangled photon pairs produced via spontaneous parametric downconversion (SPDC), utilizing frequency-resolved sum-frequency generation (SFG), the reverse process of SPDC. 
A joint phase-matching spectrum of a nonlinear crystal around 1580 nm is captured with a 40 pm resolution and a > 40 dB signal-to-noise ratio,  significantly improved compared to traditional frequency-resolved coincidence measurements. 
Moreover, our scheme is applicable to collinear degenerate sources whose characterization is difficult with previously demonstrated stimulated difference frequency generation (DFG). 
We also illustrate that the observed phase-matching function is useful for finding an optimal pump spectrum to maximize the spectral indistinguishability of SPDC photons.  
We expect that our precise spectral characterization technique will be be useful tool for characterizing and tailoring SPDC sources for a wide range of optical quantum applications.  
\end{abstract}

\maketitle

\section{Introduction}
Photons are a key resource for optical quantum-enhanced technologies, such as quantum computing, quantum communication, and quantum metrology \cite{Pan:2012kv}. 
Tailoring spectral properties of photon pairs produced by spontaneous parametric downconversion (SPDC) and spontaneous four-wave mixing (FWM)  is of great importance, since many applications require specific spectral states, in particular, with different degree of spectral entanglement; 
for example, strong frequency entanglement \cite{Tanaka:2012ua,Kaneda:2019db} is required for high-resolution quantum measurements \cite{Abouraddy:2002hm} and frequency-encoded quantum information processing \cite{Lukens:2017go}, while unentangled or indistinguishable photons \cite{Kaneda:2016fh,10.1364/oe.24.010869,Zhong:2018ez} are needed for multiplexed single-photon generation \cite{Kaneda:2017gp,Kaneda:2019bn} and quantum information processing based on multi-photon quantum interference \cite{Knill:2001is,Aaronson:2010tj}. 

High-precision engineering of spectral states of photon pairs can be achieved by high-resolution, high signal-to-noise-ratio (SNR) measurements of their spectral correlations. 
Traditionally, spectral properties of photon pairs are characterized by frequency-resolved coincidence detections \cite{10.1364/ol.30.000908,Mosley:2008ki}, where photon pairs are subjected to spectral \textit{filtering}{} followed by coincidence detections. 
However, although this approach can directly extract the joint spectral intensity (JSI) of photon pairs, its spectral resolution and SNR can be severely restricted by inefficient photon-pair generation and detection. 
Time-multiplexed techniques \cite{Avenhaus:2009hd} using group-velocity dispersion in optical fibers have demonstrated higher photon detection rates, but the required long fiber length is often accompanied by large loss that makes it difficult to achieve high-precision measurements  
(although the recent work \cite{Chen:2017gt} has achieved a relatively high spectral resolution by employing a dispersion compensation module that has a much larger dispersion and a lower loss compared to optical fibers).

To overcome the inefficiency of the coincidence detections, stimulated emission tomography (SET) \cite{Liscidini:2013cx} has been proposed for characterizing JSIs, using selective \textit{amplification}{} of spectral modes of interest via classical difference-frequency generation (DFG) and stimulated FWM processes. 
Thanks to the photon generation rates order-of-magnitude higher than SPDC and spontaneous FWM, this method has successfully demonstrated fast and high-resolution reconstruction of JSIs for various types of photon-pair sources \cite{Fang:2014dk,Rozema:2015ey,Kaneda:2016fh}. 
However, it is difficult to apply this scheme in the widely used case of collinear degenerate sources, since a seed light used for stimulated emissions becomes a source of noise photons due to its high spatial and spectral overlap with the amplified signal lights. 
Moreover, while such previous demonstrated techniques have been used for reconstructing JSI arising from the product of a pump spectral distribution and a phase-matching spectrum of a nonlinear device, an independent characterization of the two contributions would be preferable for precise diagnosis and engineering of photon pair sources.

Here we demonstrate another classical method to characterize SPDC sources, utilizing frequency-resolved sum-frequency generation (SFG), the reverse process of SPDC. 
The SFG-based characterization scheme can be applied for more general SPDC sources including collinear degenerate sources, since SFG photons can be easily separated from excitation laser lights with standard spectral filtering techniques. 
This useful nonlinear process has been demonstrated for characterizing multi-mode waveguide photon-pair sources and photonic circuits \cite{Lenzini:2018je}. 
Using the frequency-resolved SFG we demonstrate high-resolution, high-SNR measurements of a phase-matching spectrum that is purely determined by properties of a nonlinear device. 
We also show how the phase-matching spectrum obtained via the high-precision SFG measurements is useful for efficient and precise optimization of a spectral indistinguishability of SPDC photons, a key metric for optical quantum applications based on multi-photon interference. 

\begin{figure}[t!]
\centering
\includegraphics[width=1\columnwidth , clip]{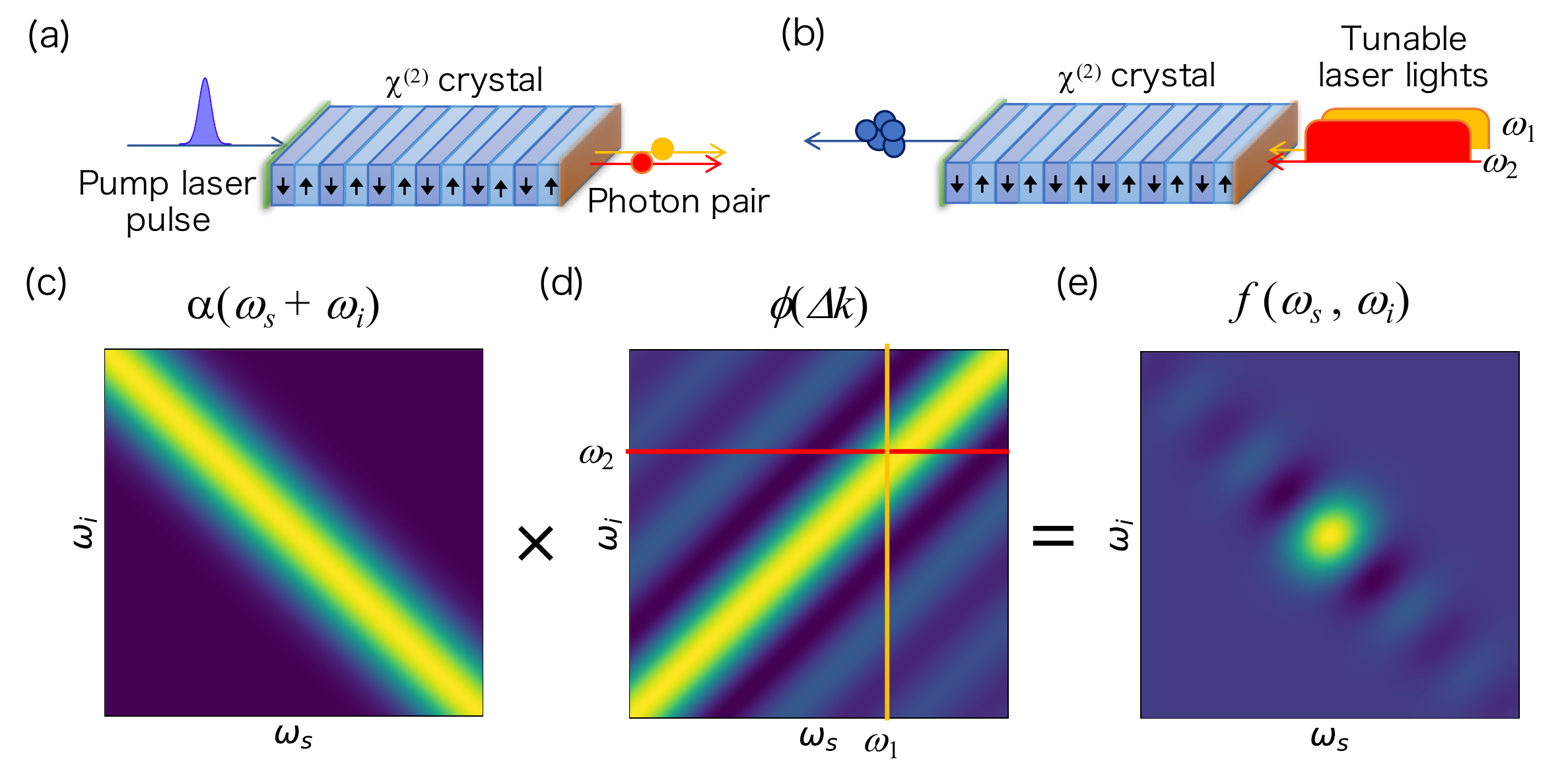}
\caption{Conceptual diagrams of (a) spontaneous parametric downconversion (SPDC) for producing photon pairs and (b) sum-frequency generation (SFG), the reverse process of SPDC that was used for our spectral characterization of an SPDC photon pair source. (c-e) Formation of (e) two-photon joint spectral amplitude (JSA) $f(\omega_s, \omega_i )$ as a product of (c) pump spectral distribution $\alpha (\omega_s + \omega_i )$ and (d) phase-matching function $\phi(\Delta k(\omega_s, \omega_i))$. The frequency-resolved SFG reveals $\phi(\Delta k(\omega_s, \omega_i))$ by scanning wavelengths of two laser sources $\omega_1$ and $\omega_2$. }
\label{SFGconcept}
\end{figure}

\section{SPDC and frequency-resolved SFG}
Conceptual diagrams of SPDC and frequency-resolved SFG, along with our scheme for spectral characterization SPDC photon pairs, are respectively illustrated in Fig. \ref{SFGconcept}(a) and (b).  
SPDC is a nonlinear optical process where one high-frequency (pump) photon is split into two low-frequency (signal and idler) photons, whereas two low-frequency photons are fused into a high-frequency photon in SFG. 
Under the plane-wave approximation for the three interacting photons, a two-photon joint spectral state produced by SPDC is given by
\begin{align}
\label{statevector}
 \ket{\psi_{si}} =  \iint d\omega_s d\omega_i  f(\omega_s, \omega_i ) \ket{\omega_s, \omega_i}, 
\end{align}
where $f (\omega_s, \omega_i )$ represents the joint spectral amplitude (JSA), and  $\ket{\omega_s, \omega_i}$ denotes a photon-pair state with signal and idler frequencies $\omega_s$ and $\omega_i$, respectively. 
Note that $ \ket{\psi_{si}}$ in Eq. \eqref{statevector} is unnormalized and $|f (\omega_s, \omega_i )|^2$ corresponds to the JSI. 
For a collinear downconversion with a pump which is not too tightly focused (and that is of our particular interest), the JSA is well approximated by 
\begin{align}
\label{JSA}
f(\omega_s, \omega_i ) \simeq  \alpha (\omega_s + \omega_i ) \phi(\Delta k(\omega_s, \omega_i)).
\end{align}
Here, $\alpha(\omega_s + \omega_i )$ is the pump spectral envelope function. 
For example, as shown in Fig. \ref{SFGconcept}(c), a pump pulse with a Gaussian spectral distribution can be described as $\alpha (\omega_s + \omega_i ) \propto \exp[(\omega_s + \omega_i -\omega_{p0} )^2/\sigma_p^2]$, where $\omega_{p0}$ and $\sigma_p$ are the pump central frequency and bandwidth, respectively. 
The phase-matching spectrum $\phi(\Delta k (\omega_s, \omega_i))$ is a function of the phase mismatch $\Delta k = k_p (\omega_p)- k_s(\omega_s) -k_i(\omega_i)$, where $k_x$ is the wavenumber of the pump ($x =p$), signal ($x =s$), and ilder ($x =i$) modes.
$\phi(\Delta k(\omega_s, \omega_i))$ is determined by a dispersion relation and a nonlinearity profile along the propagation direction of interacting photons in a nonlinear device.   
For example, the periodically-poled potassium titanyl phosphate (PPKTP) crystal \cite{Kuzucu:2005hd,Shimizu:2009uq,Evans:2010jn,10.1364/oe.19.024434, YABUNO:2012kk} to be characterized in our experiment has a uniform nonlinearity over the crystal length $L$ and satisfies a group velocity matching condition, i.e., $(k'_p-k'_s) = -(k'_p - k'_i) = D $, where $k'_x = \frac{dk_x}{d\omega_x}|_{\omega_{x0}}$ is the inverse group velocity at the photon's central frequency $\omega_{x0}$. 
In this case, the phase-matching spectrum has a sinc distribution with a positive spectral correlation: $\phi(\Delta k) = \mathrm{sinc} (D(\Omega_s-\Omega_i) L/2)$, where $\Omega_x = \omega_x -\omega_{x0}$. 
The JSA can be engineered by controlling independently the pump spectral distribution and the phase-matching spectrum of the nonlinear device, as shown in Fig. \ref{SFGconcept}(c-e).  
According to Eq. \eqref{statevector}, the SPDC photon-pair generation rate for $\omega_s$ and $\omega_i$ is $N_{\mathrm{SPDC}} (\omega_s, \omega_i) \propto N_p(\omega_s +\omega_i) |f(\omega_s, \omega_i )|^2$, where $N_p(\omega_s +\omega_i)$ is the pump photon rate at the frequency $\omega_s +\omega_i$. 

Our frequency-resolved SFG technique depicted in Fig. \ref{SFGconcept}(b) can directly reveal the phase-matching spectrum $|\phi(\Delta k)|$ without prior knowledge of the nonlinear device. 
The nonlinear device of interest is excited by two narrowband laser lights with frequencies $\omega_1$ and $\omega_2$ to produce SFG photons at $\omega_3 = \omega_1 + \omega_2$. 
SFG photons are detected by a single-photon detector, or an optical power detector. 
Since SFG is the reverse process of SPDC, its photon generation rate $N_{\mathrm{SFG}} (\omega_1, \omega_2)$ is also proportional to the same phase-matching function \cite{10.5555/1817101}:
$N_{\mathrm{SFG}} (\omega_1, \omega_2) \propto N_1(\omega_1) N_2(\omega_2) | \phi(\Delta k(\omega_1, \omega_2)) |^2$, where $N_x (\omega_x)$ ($x = 1,2$) is the input photon rate at $\omega_x$. 
Thus, one can predict the phase-matching spectrum from the distribution $N_{\mathrm{SFG}} (\omega_1, \omega_2)$ obtained by scanning the two laser frequencies. 

This SFG scheme has fundamental and technical advantages over previously demonstrated spectral measurement techniques. 
First, our SFG scheme implemented with classical measurements is fundamentally more efficient and robust against photon generation and detection inefficiencies compared to the direct measurement of SPDC photon pairs.  
Since the direct measurement scheme relies on coincidence detections of single pairs produced by single pump pulses, 
one needs to use sufficiently low pump intensity to suppress multi-pair emission (typically $\lesssim 1\%$ pair generation probability per pulse or coincidence detection time window). 
Also, the statistics of the coincidence measurement can be easily disturbed by uncorrelated detector dark counts and background photon counts. 
Meanwhile, the frequency-resolved SFG does not require coincidence detection but only single measurement of SFG photons, allowing us to produce much more than one photon per detection time window, and thus enabling a much more efficient and precise measurement in a classical way.  
Classical characterization of photon-pair sources has also been demonstrated by SET \cite{Liscidini:2013cx}, where an SPDC or spontaneous FWM process is stimulated by a seed light source that has the frequency matched to signal or idler modes.  
However, this seed source can also be a noise source when it is applied to collinear (near-)degenerate sources that are often prefered for the simplicity and the robustness of their optical setup. 
In our scheme, sum-frequency photons can be clearly separated from input lower-frequency laser lights by standard spectral filters. 
Also, the DFG scheme is strictly valid only for lossless nonlinear devices \cite{10.1364/ol.40.001460}.  
Moreover, while the previous demonstrations captured JSI, i.e., $|f (\omega_s, \omega_i )|^2$, a product of a pump spectral distribution and a phase-matching function, our frequency-resolved SFG can directly measure $|\phi(\Delta k)|$ which can be preferable for precise diagnosis of nonlinear devices (although direct measurements of $|\phi(\Delta k(\omega_s, \omega_i))|^2$ would also be possible by SET with a tunable narrowband pump source). 

\begin{figure}[t!]
\centering
\includegraphics[width=0.9\columnwidth , clip]{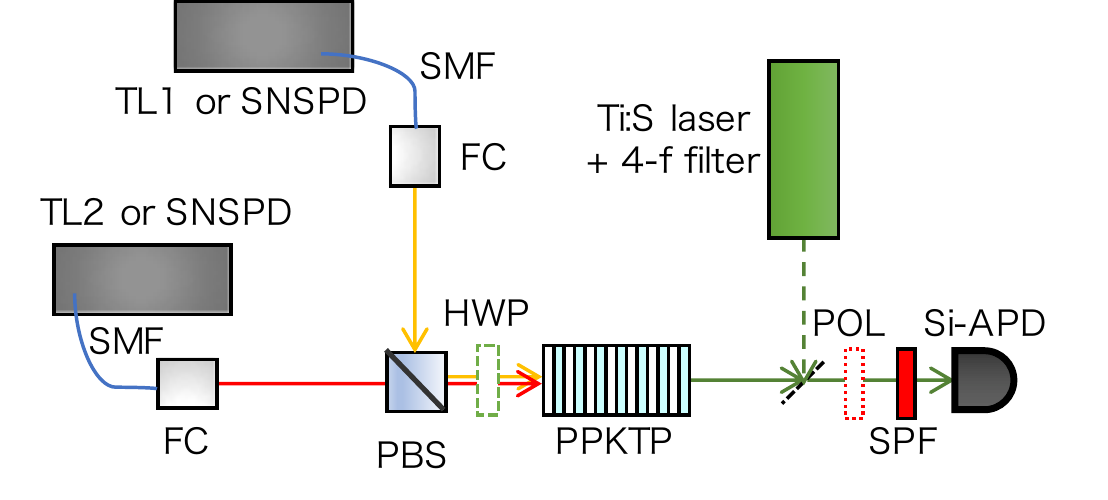}
\caption{Schematic diagram of the experimental setup. FC, fiber coupler; PBS, polarizing beamsplitter; SMF, single-mode fiber: TL1-2, tunable laser; SNSPD, superconducting nanowire single-photon detector, SPF, short-pass filter; HWP, halfwave plate; POL, polarizer; Si-APD, Si avalanche photodiode. }
\label{setup}
\end{figure}

\section{Experiment}
Our experimental setup for frequency-resolved SFG measurements is illustrated in Fig. \ref{setup}.
We applied our scheme to a periodically-poled potassium titanyl phosphate (PPKTP) crystal having a poling period $\Lambda =$ 46.1 $\mu$m with a 50\% poling duty cycle and crystal length $L = 30$ mm.  
The crystal is designed and fabricated to produce collinear degenerate photon pairs at $\sim$ 1580 nm with Type-II 1st-order ($m = 1$) quasi-phase-matching (QPM) condition, where the pump, signal, and idler photons are polarized parallel to the crystallographic Y, Z, and Y axes, respectively \cite{Konig:2004en}.  
This QPM also satisfies a group-velocity matching condition so that the spectral entanglement or indistinguishability of produced photons can be controlled by pump bandwidth \cite{Kuzucu:2005hd,Shimizu:2009uq,Evans:2010jn,10.1364/oe.19.024434, YABUNO:2012kk}; we will demonstrate that our SFG measurement is useful for efficiently finding an attainable spectral indistinguishability and an optimal pump bandwidth. 
The optical setup is first aligned using SPDC photon pairs pumped by a Ti:S laser at $\lambda_p = 790$ nm. 
Signal and idler photons are collected into different single-mode optical fibers, each of which is connected to a superconducting nanowire single-photon detector (SNSPDs) \cite{10.1364/oe.25.006796}. 
The fiber collection efficiency of SPDC photon pairs is $\sim90\%$ with pump and collection beam waists in the middle of the crystal of $300$ $\mu$m and $180$ $\mu$m. 
The collection fibers are then connected to a tunable laser (TL1 and TL2, the spectral linewidth of < 50 kHz) to perform frequency-resolved SFG measurements. 
In our experiment, SFG photons are subjected to single-photon counting measurements by a Si avalanche photodiode (Si-APD) due to the inefficient generation rates of SFG photons ($\sim$ 1 Mcps at the peak wavelengths) with the limited power of tunable laser sources ( $< 0.1$ mW); 
however, this scheme can be implemented with a standard optical detectors for sufficiently high input laser powers (> 10 mW), where the generation rate of SFG photons can be $> 10$ Gcps, corresponding to an optical power of $> 1$ nW. 

\section{Result and discussion}
\subsection{Joint phase-matching spectrum}
\begin{figure}[t!]
\centering
\includegraphics[width=1\columnwidth , clip]{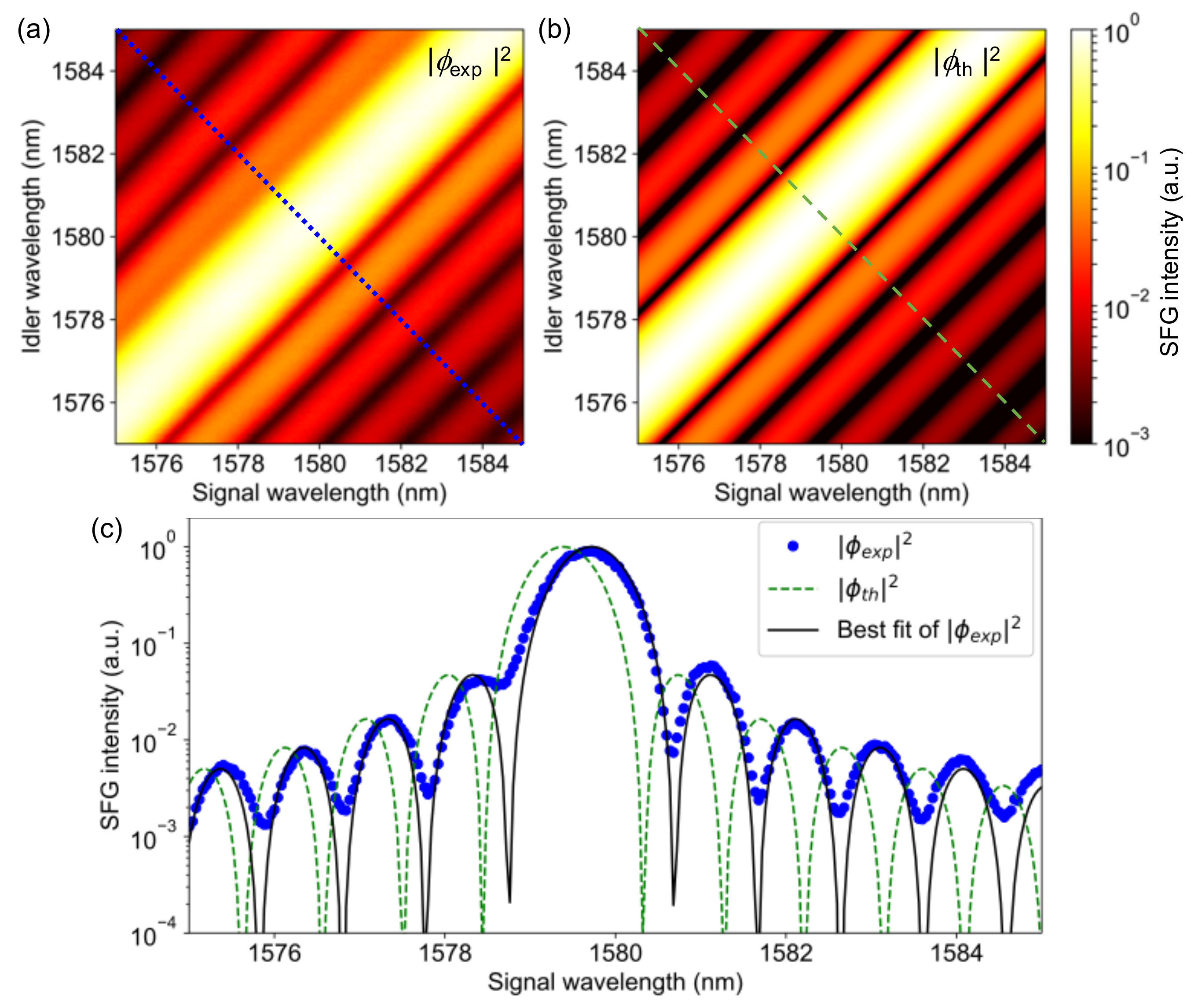}
\caption{(a) Observed phase-matching spectrum $|\phi_{\textrm{exp}}|^2$ of a Type-II PPKTP crystal via frequency-resolved SFG. (b) Theoretical prediction $|\phi_{\textrm{th}}|^2$. 
(c) Cross-sectional view of $|\phi_{\textrm{exp}}|^2$ and $|\phi_{\textrm{th}}|^2$, where the signal and idler wavelengths $\lambda_s, \lambda_i$ are anti-correlated: $1/\lambda_s = 1/\lambda_p - 1/\lambda_i$, and $\lambda_p = 790$ nm. }
\label{SFG}
\end{figure}

Figure \ref{SFG}(a) shows our observed phase-matching function $|\phi_{\textrm{exp}}|^2$ via frequency-resolved SFG. 
The signal and idler wavelength were scanned in steps of 0.04 nm across a 10 nm bandwidth.  
Our measurement SNR, the ratio of the peak count rate ($> 1$ Mcps) by the dark count rate of our Si-APD ($<$ 100 cps), is $> 40$ dB, largely improved over the methods using coincidence detections (10-30 dB) \cite{Evans:2010jn,YABUNO:2012kk,Chen:2017gt}. 
Note again that it is difficult to characterize this collinear degenerate source by the SET scheme. 
The data acquisition time in our measurement is approximately 20 hours due to the sequential $251^2 = 63001$ measurements with the 1-second interval and the limited laser power ($< 0.1$ mW). 
However, the measurement can be made much more efficient by employing higher-power lasers and optical power detectors.
We also note that the efficiency of our scheme could be further improved by implementing spectral-multiplexed measurements replacing one of tunable narrowband lasers and a single-pixel detector with a broadband (white) light source and a multi-pixel spectrometer. 
We then compare $|\phi_{\textrm{exp}}|^2$ with the theoretical prediction $|\phi_{\textrm{th}}|^2$ shown in Fig. \ref{SFG}(b) taking into account for the Sellmeier equations \cite{Konig:2004en} and the crystal fabrication parameters $\Lambda$ and $L$. 
To visualize the difference, we also show Fig. \ref{SFG}(c), cross sections of Fig. \ref{SFG}(a,b), where the signal and idler wavelengths $\lambda_s, \lambda_i$ are anti-correlated: $1/\lambda_s = 1/\lambda_p - 1/\lambda_i$, and $\lambda_p = 790$ nm. 
Our observed sinc-like distribution with positive spectral correlation originated from the uniform nonlinearity and the group-velocity-matching condition in the PPKTP crystal, is in reasonable agreement with $|\phi_{\textrm{th}}|^2$. 
However, the observed peak wavelengths and the period of the sinc peripheral lobes are slightly different from our theoretical predictions, as shown in Fig. 3(c). 
With the best fit of the experimental data shown as the solid curve in Fig. \ref{SFG}(c), we obtained $\Lambda =$ 46.125 $\mu$m and $L = 29.0$ mm: note that although we used the fixed Sellmeier equations \cite{Konig:2004en} for this data fitting, we will see in next section that suggests that the Sellmeier equations also need modifications. 
We also see that the contrast between peaks and dips in the observed spectrum is somewhat degraded from that of the theoretical sync distribution. 
This might be due to the fluctuation of the poling period. 
Further investigation of the phase-matching spectrum is out of the scope of the paper and will be discussed elsewhere. 
Nonetheless, our high-precision SFG measurement has successfully resolved the slight difference between the designed and actual phase-matching spectra. 

\subsection{Wideband second harmonic generation}
We then applied our scheme to characterize the details of strictly degenerate SPDC in the PPKTP crystal. 
For this experiment, a single tunable laser light with a diagonal polarization, i.e., a superposition state of Y and Z polarizations, is used as a fundamental light to produce Type-II second-harmonic generation (SHG). 
Compared to the two-dimensional $(\omega_1,\omega_2)$ SFG measurements, this simplified one-dimensional $(\omega_1)$ measurement can efficiently investigate the phase-matching distribution for a wide spectral range. 

\begin{figure}[t!]
\centering
\includegraphics[width=0.9\columnwidth , clip]{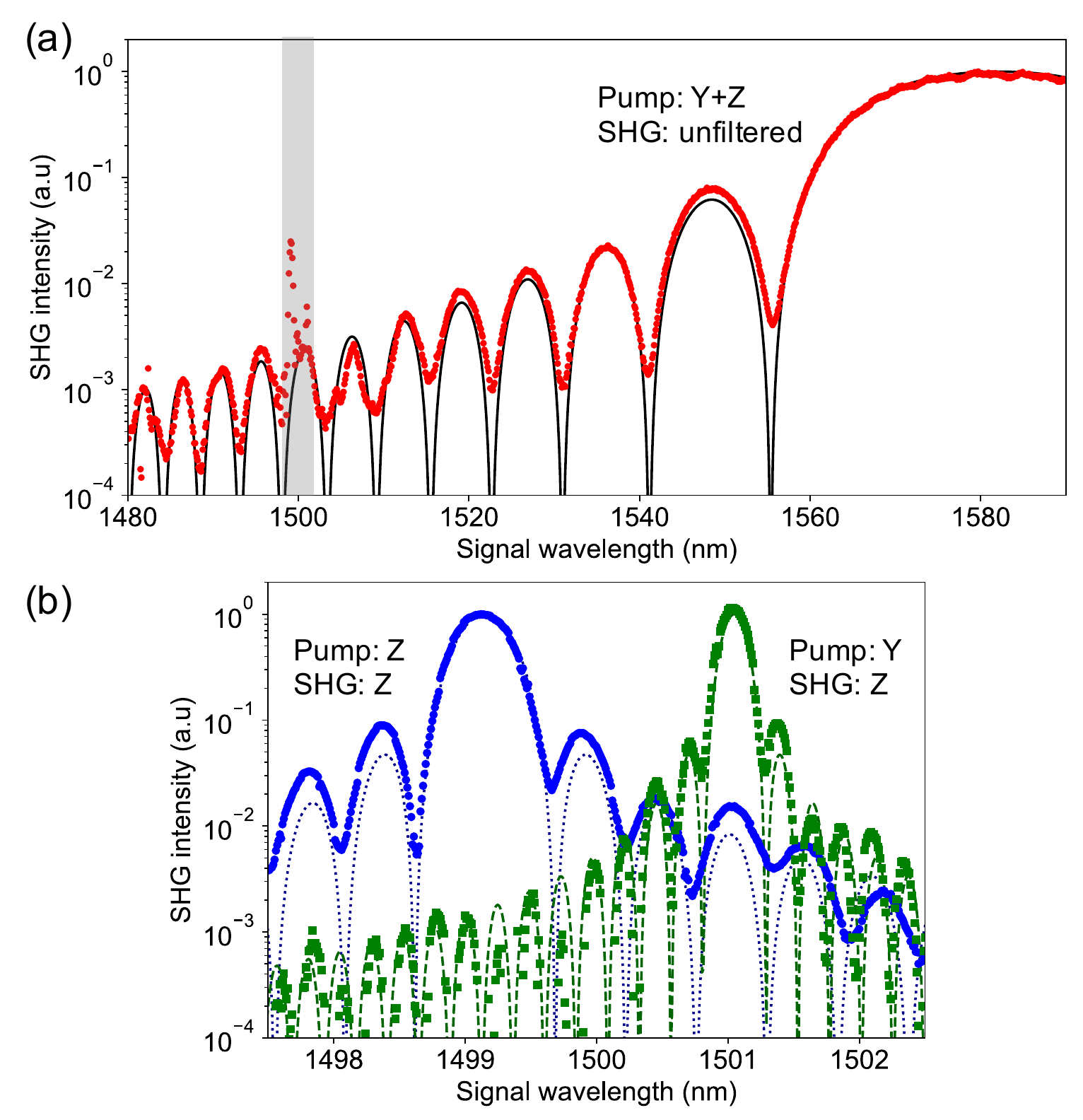}
\caption{Observed SHG spectra for the PPKTP crystal. (a) Wideband spectrum without filtering the polarization of SHG photons. The shaded area is where we find unexpected Type-I and Type-0 QPM peaks. (b) Measured SHG around 1500 nm for different fundamental and SHG polarizations. The solid, dashed, and dotted curves are the best fit of the phase-matching spectra with the 1st-order Type-II, 7th-order Type-I, and 2nd-order Type-0 QPM functions, respectively. }
\label{SHG}
\end{figure}

Figure \ref{SHG}(a) shows our observed SHG intensity versus the fundamental laser wavelength. 
Our measurement scanned across the laser tuning range (1480-1590 nm) has successfully revealed 11 peripheral lobes in which the minimum peak intensity is only 0.1\% of the maximum peak around 1580 nm.
Note that, due to the group-velocity matching condition, the main peak has a considerably wide bandwidth compared to those in normal phase-matching conditions. 
In most of the spectral region, the observed spectrum closely matches the theoretical predictions taking into account the $L$ and $\Lambda$ obtained via the best fit of the SFG spectrum shown in Fig. \ref{SFG}(c). 
Meanwhile, we also find unexpected peaks around 1500 nm that are far off the main peak at 1580 nm and thereby difficult to be detected with ordinary narrowband spectral measurements. 
By performing polarization-resolved SHG measurements using a half-wave plate and a polarizer (see Fig. \ref{setup}), we identified that there are two peaks produced by Type-I ($\textrm{Y}+ \textrm{Y} \rightarrow \textrm{Z}$) and Type-0 ($\textrm{Z}+ \textrm{Z}\rightarrow \textrm{Z}$) polarization configurations, as shown in Fig. \ref{SHG}(b). 
With the best fit of the Type-I(0) peaks we obtained the QPM order $m = 7 (2)$ and the poling period $\Lambda =  45.807$ $\mu$m (46.010 $\mu$m). 
The slight difference in the estimated poling periods for the three different QPM conditions suggests that the Sellmeier equations need slight modifications: the detailed investigation will be discussed elsewhere. 
Moreover, the observed non-zero 2nd-order QPM SHG indicates that the duty cycle of our crystal is not strictly matched to the designed value (50\%), where an even-order QPM condition has effective zero nonlinearity while the 1st-order QPM has a maximum efficiency.

\subsection{Optimization of spectral indistinguishability}
\begin{figure}[t!]
\centering
\includegraphics[width=1\columnwidth , clip]{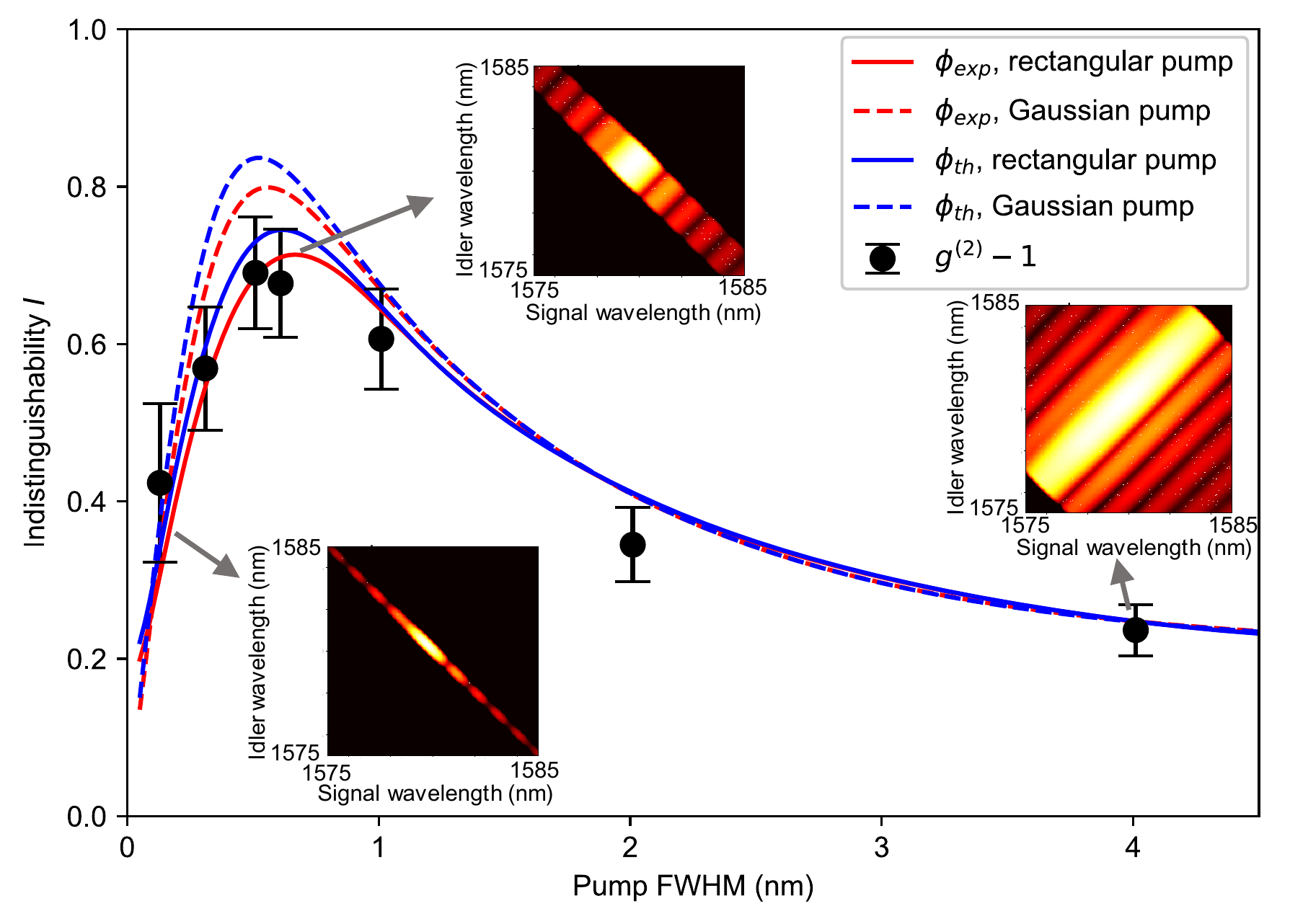}
\caption{Spectral indistinguishability of SPDC photons estimated by the Schmidt decomposition \cite{10.1134/s1054660x06060041} of the simulated JSAs. 
The solid (dashed) line is for the observed (theoretical) phase-matching spectrum $\phi_{\textrm{exp}}$ ($\phi_{\textrm{th}}$). The red (blue) line is for the Gaussian (rectangular) pump spectral distribution. For comparison, we also show the measured indistinguishability using $g^{(2)}$ measurements ($I = g^{(2)} -1$) with the rectangular pump spectrum (black circles). Insets show JSIs for different rectangular pump bandwidths and $\phi_{\textrm{exp}}$ that are investigated by the $g^{(2)}$ measurements.}
\label{I}
\end{figure}

Finally, we demonstrate that our SFG measurement is useful for optimizing a spectral indistinguishability of individual SPDC photons. 
Single-photon indistinguishability is a key metric for many optical quantum applications utilizing multi-photon interference. 
In order to maximize the indistinguishability for a given nonlinear device and its phase-matching spectrum, one needs to optimize a pump spectral distribution, i.e., another contribution to the JSA, as shown in Eq. \eqref{statevector}. 
However, direct measurement schemes of the indistinguishability such as Hong-Ou-Mandel two-photon interference \cite{Hong:1987vi, Mosley:2008ki} and the second-order autocorrelation function \cite{Christ:2011ku} need to acquire inefficient multi-pair generation events that can hamper high-precision measurement and optimization. 
Here we demonstrate that the phase-matching spectrum $|\phi(\Delta k (\omega_s,\omega_i))|$ obtained by high-precision SFG measurement is useful  for precise optimization of the indistinguishability by simulating the JSA with a numerical pump spectral distribution. 
Figure \ref{I} shows the simulated spectral indistinguishability obtained by the Schmidt decomposition \cite{10.1134/s1054660x06060041} of the numerically predicted JSAs using the PPKTP crystal. 
We simulated JSAs for the observed and theoretical phase-matching function $\phi_{\textrm{exp}}$ and $\phi_{\textrm{th}}$ shown in Fig. \ref{SFG}(a) and (b) (with the assumption that $\phi_{\textrm{exp}}$ has no imaginary part).
As a pump source, we assumed Gaussian and rectangular spectral distributions: 
the former is a good approximation for many pulsed lasers, while the latter is close to the spectral distribution of our SPDC pump source that is used for the measurement of the second-order autocorrelation function, as will be discussed later. 
The attainable spectral indistinguishabilities with $\phi_{\textrm{exp}}$ ($I =$ 80\% and 71\% for Gaussian and rectangular pump) are slightly lower than the ones with $\phi_{\textrm{th}}$ ($I =$ 84\% and 75\% for the Gaussian and rectangular pump). 
This may be due to the lower contrast in $\phi_{\textrm{exp}}$, caused by the slight fabrication errors found in our SFG and SHG measurements. 

For comparison, we also estimated the indistinguishabilty by the measurement of the second-order autocorrelation function $g^{(2)}$ \cite{Christ:2011ku}, shown as black circles in Fig. \ref{I}. 
In our $g^{(2)}$ measurement, the PPKTP crystal was pumped by broadband Ti:S laser pulses passed through a 4-f spectral filter, where a pump pulse spectrum is shaped into an approximate rectangular form with a tunable bandwidth. 
Produced SPDC signal photons coupled into a single-mode fiber are measured by a fiber-based Hanbury-Brown-Twiss setup \cite{Brown:1956vw}. 
We see that the indistinguishabilities measured by the $g^{(2)}$ measurements ($I = g^{(2)} -1$) are in excellent agreement with our simulation for  
the observed phase-matching spectrum and rectangular pump function. 
However, the indistinguishabilities estimated from the $g^{(2)}$ measurements have large statistical uncertainties due to inefficient double-pair emission probability ($\sim 10^{-5}$ per pulse) and the resulting coincidence detection rate of only $\sim$ 1000 for 30 minutes. 
Thus, our method based on the SFG measurement has an ability to optimize precisely the pump spectral bandwidth for maximizing the spectral indistinguishability. 
We also note that the $g^{(2)}$ measurement scheme can easily overestimate the indistinguishability by imperfect separation of signal and idler photons and thus needs careful construction of the optical setup \cite{10.1364/oe.19.024434} (see Appendix for details).  

\section{Conclusion}
We have demonstrated how frequency-resolved SFG, the reverse process of SPDC, can be used for characterizing spectral properties of SPDC photons. 
Our scheme implemented with efficient classical measurement has successfully captured the phase-matching spectrum of a PPKTP crystal with a 0.04-nm resolution and a 40-dB SNR. 
This high-precision measurement unaffected by an SPDC pump source is useful for the diagnosis of nonlinear devices as well as efficient optimization of the source performance. 
Our scheme can be applied to a variety of SPDC sources, including collinear degenerate sources thanks to the large spectral separation of SFG photons and the pump laser lights. 
Introducing spectral-multiplexed measurements and higher-power lasers, the data acquisition time will be significantly improved.  
We expect that this high-resolution, high-SNR measurement technique will be useful tool for characterizing and tailoring SPDC sources for a wide range of optical quantum applications.

\section*{Appendix: effect of the imperfect separation of signal and idler photons in $g^2$ and SFG measurements}
We find that the $g^{(2)}$ measurement scheme can easily overestimate the indistinguishability due to an imperfect separation of signal and idler photons and their coupling to the same collection fiber. 
When the error probability $R_e$ for the separation of signal and idler photons (e.g., the reciprocal of the PBS's extinction ratio in our experiment) is comparable or larger than the SPDC pair-generation probability $p$, the unwanted signal-idler pair coincidence probability $\sim p R_e$ is not negligible compared to the signal-signal coincidence probability $\sim p^2$ due to double-pair emissions. 
This is in particular critical for collinear degenerate sources, where photon pairs are initially overlapped in spectral, temporal, and spatial modes; in our preliminary $g^{(2)}$ measurement, we observed $g^{(2)} > 4$ due to a misaligned PBS with a degraded extinction ratio ($R_e = 0.01$). 

The corresponding error in the SFG measurement would be the pump preparation error, for example, a laser light prepared for exciting the signal polarization mode which would also have a small overlap to the idler polarization. 
However, such error signals can be produced when two laser lights exciting signal and idler modes \textit{both} receive the state preparation errors with the probability $R_e^{2}$, much less than that of the peak SHG signal. 
Thus, although this SFG error signal probability is not directly related to the estimation error for the indistinguishability (since it also depends on pump spectral distribution), our frequency-resolved SFG is technically more robust. 

\section*{Funding}
JSPS KAKENHI Grant Number JP18H05949 and JP19H01815, MEXT Quantum Leap Flagship Program (MEXT Q-LEAP) Grant Number JPMXS0118067581, and Matsuo Academic Foundation.

\section*{Acknowledgments}
We thank So-young Baek and Pierre Vidil for helpful discussions. 


\bibliography{export_2020-6-25}

\end{document}